\documentstyle[preprint,aps,eqsecnum]{revtex}
 
\def\beq#1{\begin{equation} \label{#1}}
\def\eeq{\end{equation}}

\def\ket#1{\left\vert #1\right\rangle}

\input prepictex
\input pictex
\input postpictex
\newdimen\tdim
\tdim=\unitlength
\def\stpltsmbl{\setplotsymbol ({\small .})}

\newbox\phru
\setbox\phru=\hbox{\beginpicture
\setcoordinatesystem units <\tdim,\tdim>
\stpltsmbl
\setquadratic
\plot
0 0
2.5 3
5 0
7.5 -3
10 0
/
\endpicture}
\def\photonru #1 #2 *#3 /{\multiput {\copy\phru}  at
#1 #2 *#3 10 0 /}
 
\newbox\sru
\setbox\sru=\hbox{\beginpicture
\setcoordinatesystem units <\tdim,\tdim>
\stpltsmbl
\setquadratic
\plot
  0.0   0.0
  4.8   1.5
  7.5   5.0
  7.3   8.5
  5.0  10.0
  2.7   8.5
  2.5   5.0
  5.2   1.5
 10.0   0.0
/
\endpicture}
\def\springru #1 #2 *#3 /{\multiput {\copy\sru}  at
#1 #2 *#3 10 0 /}

\begin{document}
{
\tighten
 
\title{Experimental Puzzles in Heavy Flavor Decays\\
Anomalously high $\eta'$ appearance in
charmless strange  $B$ decays - Flavor SU(3) breaking in Charm Decays}

\author{Harry J. Lipkin\,\thanks{Supported 
in part by grant from US-Israel Bi-National Science Foundation
and by the U.S. Department
of Energy, Division of High Energy Physics, Contract W-31-109-ENG-38.}}
\address{ \vbox{\vskip 0.truecm}
  Department of Particle Physics
  Weizmann Institute of Science, Rehovot 76100, Israel \\
\vbox{\vskip 0.truecm}
School of Physics and Astronomy,
Raymond and Beverly Sackler Faculty of Exact Sciences,
Tel Aviv University, Tel Aviv, Israel  \\
\vbox{\vskip 0.truecm}
High Energy Physics Division, Argonne National Laboratory,
Argonne, IL 60439-4815, USA\\
~\\harry.lipkin@weizmann.ac.il
\\~\\
}

\maketitle
 
\begin{abstract}

Simple experimental tests are proposed which can clarify the origin for the
anomalously high $\eta'$ appearance in charmless strange final states in $B$
decays and can investigate the the nature of SU(3) symmetry-breaking in weak
heavy flavor decays.. 

\end{abstract}
} 

\renewcommand{\baselinestretch}{1.2}
\setlength{\baselineskip}{16pt}

\pagebreak

Most theoretical predictions in heavy flavor physics begin with well defined 
models and assumptions.  When the predictions disagree with experiment,
possibly indicating the existence of interesting overlooked physics, it is not
clear which of the assumptions  underlying the predictions have gone wrong
or whether there may be clues to evidence for new physics beyond the 
standard model. This is particularly relevant to CP violation where the 
evidence supporting the Kobayashi-Maskawa phase as the explanation is 
essentially one piece of data fit by one free parameter and any unexpected
experimental signals should be thoroughly explored. 

We develop an alternative approach to examine two puzzling phenomena where
experimental results challenge conventional wisdom:  (1) the anomalously high
$\eta'$ appearance in charmless strange final state in $B$ decays ; (2) Some
apparent violation of SU(3) predictions relating Cabibbo-favored and
doubly-forbidden transitions in charm decays. We look for  experimental tests 
which can bring new insight  into apparent contradictions and hopefully bring
evidence for new physics that may be hidden in the puzzles.

\section{The $B\rightarrow K\eta$ - $B\rightarrow K\eta'$ problem} 

The large experimental branching ratio\cite{PDG} 
$BR(B^+ \rightarrow  K^+ \eta')=6.5\pm 1.7 \times 10^{-5}$ as compared with
$BR(B^+ \rightarrow  K^+ \eta) < 1.4 \times 10^{-5}$ 
and $BR(B^+ \rightarrow  K^o \pi^+)=2.3\pm 1.1 \times 10^{-5}$ 
still has no completely satisfactory explanation and has aroused considerable
controversy\cite{PHAWAII}.

 \subsection {A parity selection rule can separate two types of models }

We note here a clear experimental method to distinguish between  two
mechanisms  proposed to explain this  high $\eta'$ appearance  and the high
$\eta'/\eta$ ratio in charmless strange $B$ decays. 

1. Treatments where the enhancement arises from an additional diagram; e.g. the
anomaly, gluon couplings to the flavor singlet component of the $\eta'$ or
intrinsic charm\cite{HALZHIT,ATSON}. This diagram is often called an
``OZI-forbidden hairpin 
diagram" and is described by fig. 1 using the flavor-topology
description\cite{bkpfsi}.  The  enhancement is universal and should appear in
all similar final states. In particular, it should be independent of the
$parity$ of the final state.

2. Treatments where the enhancement arises from constructive interference
between diagrams producing the $\eta'$ via the strange and nonstrange 
components of the $\eta'$ wave function\cite{bkpfsi}. The sign of the
interference should be constructive only for even parity final states and
destructive for odd parity final states; and vice versa for the corresponding
interference in the $\eta$ wave function.  One example is the model where the
gluonic penguin diagram produces the $\eta$ and $\eta'$  in charmless strange
final states both via the $u \bar u$ (fig.2) (or $d \bar d$) and $s \bar s$
(fig.3) components of these mesons, denoted respectively as $\eta_u$ and
$\eta_s$.  The two components interfere constructively for the $\eta'$  and
destructively for the $\eta$ in all final states of even parity and vice versa
for states of odd parity. 

This model predicts a parity selection rule in which  the $\eta/\eta'$ ratio 
should be large in charmless strange final states  of ODD parity and small in
states of EVEN parity. This selection rule should be violated in models of type
1 above which introduce some other parity-independent mechanism for explaining
the large $\eta'$  enhancement found in the $B\rightarrow K\eta'$ 
decay\cite{HALZHIT,ATSON} 

That these considerations lead to a large $\eta'$/$\eta$ ratio for the $K\eta$
and $K\eta'$  final states and the reverse for the $K^*$(892) $\eta$ and $K^*$
$\eta'$  has been pointed out\cite{PHAWAII}. This seems to agree with
experiment, although so far  the $K^*$ $\eta$ has been seen and the $K^*$
$\eta'$ has not. 

Note that the simple tree diagrams (figs. 4 and 5) can produce the $\eta$ and
$\eta'$ only via their nonstrange components. These contributions are
eexpected to be relatively small and in any case cannot contribute to a large
$\eta'$/$\eta$ ratio since these diagrams contribute roughly equally to both
final states.

\subsection {Experimental consequences of the Parity Selection Rules}

We now note some further experimental consequences of this
parity rule which can be checked possibly with already available data.

\begin{enumerate}

\item The $K\pi\eta$ and $K\pi\eta'$ states all have odd parity, even when the
$K\pi$ is not in a $K^*$. Therefore the selection rule predicts that the
$K\pi\eta$ should be much stronger than $K\pi\eta'$ when summed over all final
states.  Using the  better statistics  obtainable by summing over all  charged
and neutral $B$ decays. one may get a clear  test between the two models.  A
strong enhancement of the $\eta$ over the $\eta'$  would provide  strong
evidence against models that produce the $\eta'$  via the SU(3) singlet
component; e,g, gluons, anomaly or intrinsic charm.

\item An appreciable inclusive signal for $B \rightarrow K \eta' X$ has been
reported. A measurement of the spectrum of the ``missing mass" $M_X$ can
check the validity of the parity selection rule, which requires $M_X$ to be
at least two pion masses. Confirmation of this selection rule would also
simplfy the partial wave data analysis by ruling out large contributions from
resonances with large $K\pi$ decay modes  without needing any complicated
fits to mass plots;  e.g. the scalar $K_o(1430)-(93\%-K\pi)$ , the tensor
$K_2(1430)-(50\% -K\pi)$ and  higher respnances like 
$K^*(1680)-(39\%-K\pi)$.

\item The measurement of the TRANSVERSITY in the final states
 $\eta \rho K$,
$\eta' \rho K$, $\eta \pi K^*(892)$ and $\eta' \pi K^*(892)$ 
 gives an
unambiguous signal for the PARITY of the final state (whether it is $0^+$ or
$0^-$) independent of the quantum numbers of the $K\pi\pi$ state recoiling
against the $\eta$ or $\eta'$. This is the
measurement of the polarization of the vector meson in its rest frame with
respect to an axis normal to the VPP plane\cite{Trans} 
 
\item An $\eta$ or $\eta'$  recoiling against a $K^*$ resonance with NATURAL
parity (even P for even J and odd P for odd J) has odd parity and should give a
final state favoring the $\eta$ over the $\eta'$ . The opposite is true for a
recoil against a state with UNNATURAL parity. the $K$ and $K^*(892)$ states are
special cases of this prediction. 
 
\item One should look for $K\eta$ and $K\eta'$ resonances in the states
$K\eta X$ and $K\eta' X$. Here the even parity resonances should favor the
$\eta'$  and the odd parity resonances favor the $\eta$. 
  
\end{enumerate}

\subsection {Possible new physics and CP violation}

If the parity selection rule is violated, there is always a possibility that
it is due to new physics that can produce CP violation. One simple test of
any far-out idea for direct CP-violation is to compare corresponding $B^+$
and $B^-$ decays and look for a difference. Since a number of decays to
final  states containing the $\eta'$ seem to be enhanced considerably beyond
what is expected from conventional models, it would seem reasonable and cheap
to check for direct CP violation in such cases.  Other cases of anomalously
large decays to $\eta'$ final states which might merit special investigation
include $D_s \rightarrow \pi \eta'$ and $D_s \rightarrow \rho \eta'$. 

\section {Experimental Puzzles in Doubly-Cabibbo Suppressed Charm Decays}

Cabibbo-favored and doubly-cabibbo suppressed charm decays have been noted to
go into one another\cite{plipzh,lincoln} under an SU(3) transformation which
interchanges $d$ and $s$ flavors everywhere, This transformation is a subgroup of SU(3)
sometimes called a Weyl reflection or a  U-spin reflection. 

\beq{QQ3}       
d \leftrightarrow s  ; ~ ~ ~    K^+ \leftrightarrow \pi^+  ; ~ ~ ~
  K^- \leftrightarrow \pi^-  ; ~ ~ ~   D^+ \leftrightarrow D_s    
\eeq

 Two aspects of this relation which suggest interesting implications of any
 symmetry breaking are relevant here.

  (1)  Experimental tests of the magnitude of SU(3) breaking will be relevant
in the interpretation of information about the CKM matrix and the unitarity
triangle obtained from. standard model analyses of weak decays which assume 
SU(3) symmetry. 

  (2) In the standard model the Cabibbo-favored and doubly-suppressed charm
decays are proportional to the same combinations of CKM matrix elements and
no direct CP violation can be observed. Thus any evidence for new physics that
can introduce a CP-violating phase between these to amplitudes deserves serious 
consideration\cite{plipzh}.   
  
\subsection {Relations between Cabibbo-Favored and Doubly-Cabibbo Suppressed
$D^o$ decays}.

Wolfenstein\cite{lincoln} has noted that the $D^o (c \bar u)$ which contains no
$d$ nor $s$ quarks is invariant under this $d-s$-interchange SU(3) 
transformation (\ref{QQ3}) and that under this transformation   the $ K^+ \pi^-
$ and $ K^- \pi^+ $ decay modes go into one another as seen in figs. 6 and 7. 
Thus SU(3) predicts that the doubly-Cabibbo-suppressed and Cabibbo-favored
decays to these final states should have the same strong phases. 

A recent analysis of the two-pseudoscalar decay modes of the neutral $D$
mesons\cite{sven} suggests that the phases are not the same. But the $ K^+
\pi^- $ and $ K^- \pi^+ $ final states are charge conjugates of one another.
Thus a strong phase difference cannot be introduced by any $K-\pi$ final state
rescattering mechanism that conserves charge conjugation; e.g. Regge exchange
models\cite {GW,fsi1,fsi2,fsi3,Falk}, even if SU(3) is broken. SU(3) can be
broken in strong interactions without breaking charge conjugation only in the
hadronization transition from the quark level to the hadron level.
 \beq{hadquark1} D^o(c\bar u)
\rightarrow  (su\bar d) \bar u \rightarrow  (s \bar u\rightarrow K^{*-} )_S
\cdot  (u\bar d\rightarrow  M^+ )_W \rightarrow  K^{*-} M^+  \rightarrow
K^-\pi^+ \eeq 
 \beq{hadquark2} D^o(c\bar u) \rightarrow  (du\bar s) \bar u
\rightarrow  (d \bar u \rightarrow M^-)_S \cdot (u\bar s \rightarrow K^{*+} )_W
\rightarrow  M^-  K^{*+}  \rightarrow \pi^- K^+ \eeq  
where $K^{*\pm}$ can
denote a kaon or any $K^*$ resonance and $M^{\pm}$ can  denote a pion or any
charged meson resonance, and the subscripts S nd W  denote strong and weak form
factors. The quark-antiquark pair created from the the weak vertex is expected
to hadronize with a weak pointlike form factor and the pair including the
spectator is expected to hadronize with a hadronic form factor.

It is also of interest to examine corresponding $D$ decays with other charges 
\beq{hadquark0} D^o(c\bar u)
\rightarrow  (su\bar d) \bar u \rightarrow (u\bar u\rightarrow  M^o )_S
\cdot  (s \bar d\rightarrow \bar K^{o} )_W \rightarrow   \bar K^{o} M^o  
\rightarrow \bar K^{o}\pi^o  \rightarrow  K_S\pi^o \eeq 
 \beq{hadquark02} D^o(c\bar u)
\rightarrow  (du\bar s) \bar u
\rightarrow (u\bar u\rightarrow  M^o )_S
\cdot  (d \bar s\rightarrow  K^{o} )_W \rightarrow   K^{o} M^o  
\rightarrow  K^{o}\pi^o  \rightarrow  K_S\pi^o \eeq

\beq{axialc2}
D^+(c\bar d) \rightarrow  (su\bar d) \bar d
\rightarrow  (s \bar d\rightarrow \bar K^{o} )_S \cdot 
(u\bar d\rightarrow  M^+ )_W
\rightarrow \bar K^o M^+ \rightarrow \bar K^o \pi+ \rightarrow  K_S\pi^+
\eeq 

\beq{axiald2}
D^+(c\bar d) \rightarrow (du\bar s) \bar d
\rightarrow  (d\bar s\rightarrow  K^{o} )_S \cdot 
(u\bar d\rightarrow  M^ + )_W
\rightarrow  K^o M^+ \rightarrow  K^o\pi^+ \rightarrow  K_S\pi^+
\eeq 
Where we note that the cabibbo-favored and doubly-suppressed amplitudes leading
to final states with neutral kaons can interfere when the kaons are detected
as $K_S$. This is of particular interest if there is any CP-violating new
physics contribution. 

The entire cascade of transitions (\ref{hadquark1}) goes
into (\ref{hadquark2}) under the SU(3) $s \leftrightarrow d$ transformation,
and all the purely hadronic transitions also under charge conjugation. 
Therefore the only strong interaction mechanisms which can break SU(3) without
breaking charge conjugation must occur at the quark level and involve the 
differences between strange and nonstrange weak and strong form factors.

For the case where the two quark-antiquark pairs hadronize directly into the 
$K^+ \pi^- $ and $ K^- \pi^+ $ final states the only SU(3) breaking effect 
beyond the CKM matrix elements is the weak form factor which intoduces a 
factor of $f_\pi$ into the transition (\ref{hadquark1}) and a factor of $f_K$
into (\ref{hadquark2}). This can change the relative magnitudes of the two 
amplitudes but does not easily introduce a phase difference. The strong form
factors involve overlap integrals between different ground state s-waves and
cannot  easily introduce a phase. 

We are then led to look for transitions where the 
two quark-antiquark pairs hadronize first into another intermediate state
and then scatter by $C$-invariant strong interactions into the 
$ K^+ \pi^- $ and $ K^- \pi^+ $ final states. 
Leading candidates are the Cabibbo allowed decay modes 
$K^- a_1(1260)^+$ and $K^*(892)^-\rho^+$ observed experimentally with 
higher branching ratios than $K^-\pi^+$. 
\beq{DKBR}
BR[D^o \rightarrow  K^- a_1(1260)^+] = 7.3 \pm 1.1 \% ;~ ~ ~ 
BR[D^o \rightarrow  K^*(892)^-\rho^+] = 6.1 \pm 2.4\%
\eeq 
\beq{DKpiBR}
BR[ D^o \rightarrow  \bar K^o a_1(1260)^o] < 1.9 \% ;~ ~ ~
BR(D^o \rightarrow  K^-\pi^+)  = 3.83 \pm 0.09\%.
\eeq 
The correspondinng 
doubly-Cabibbo-suppressed  decay modes are therefore predicted by SU(3) to
have higher branching ratios than the observed DCSD 
$ D^o \rightarrow K^+ \pi^- $. 
Furthermore, since they are both 
positive parity states, like $K^-\pi^+$, all three states are coupled 
together by final state rescattering. 
Since the $a_1$ and the pion have very different wave functions and are not
related at all in the SU(3) limit, one might expect a difference both in 
magnitude and phase between the transitions
\beq{axial1}
D^o(c\bar u) \rightarrow  (su\bar d) \bar u
\rightarrow  (s \bar u\rightarrow K^{-} )_S \cdot 
(u\bar d\rightarrow  a_1^+ )_W
\rightarrow  K^{-} a_1^+ 
\rightarrow K^-\pi^+
\eeq 
\beq{axial2}
D^o(c\bar u) \rightarrow  (du\bar s) \bar u
\rightarrow  (d \bar u \rightarrow a_1^-)_S \cdot 
(u\bar s \rightarrow K^{+} )_W
\rightarrow  a_1^-  K^{+ } 
\rightarrow \pi^- K^+
\eeq 
The suggestion of a large difference is reinforced by noting 
the large difference between the experimental branching ratios
for the charged final state (\ref{DKBR}) and the neutral (\ref{DKpiBR}) for
which the  transition is described as
\beq{axial0}
D^o(c\bar u) 
\rightarrow  (su\bar d) \bar u \rightarrow (u\bar u\rightarrow  a_1^o )_S
\cdot  (s \bar d\rightarrow \bar K^{o} )_W \rightarrow   \bar K^{o} a_1^o  
\rightarrow \bar K^{o}\pi^o  \eeq 

The data seem to indicate that the product of a weak axial form factor and a
strong kaon form factor is very different from the reverse product.  This is
also expected in any model which uses factorization for the weak transition.
The data for all $D$ and $B$ decays indicate 
that decays to final states containing
the charged $a_1$ are consistently much stronger than decays to final states in
the same isospin  multiplet containing the neutral  $a_1$. The data systematics 
seem to suggest a kind of ``vector dominance" model in which $D$ and $B$ decays
to quasi-two-body final states are dominated by diagrams in which a charged $W$
boson turns into a charged pseudoscalar, vector or axial-vector meaon. However,
better data are needed to clarify this issue. 

We now note that the rescatterings 
$ K^{-} a_1^+ \rightarrow K^-\pi^+$ and $ K^{-} a_1^+ \rightarrow \bar K^o\pi^o$
can proceed by the same $\rho$ exchange mechanism that has been used in the 
Regge exchange models\cite {GW,fsi1,fsi2,fsi3,Falk} for $K-\pi$ final state 
rescattering. Since the strong $\rho \pi a_1$ coupling is comparable to 
$\rho \pi \pi$ it seems natural to extend these models to include the
the $ K^{-} a_1^+ \rightarrow \bar K \pi $ transition in addition to the 
$K-\pi$ elastic and charge exchange scattering. 

Since the charged D decays (\ref{axialc2}) go to I=3/2 final states which are
exotic and have no resonances, it might be a reasonable first approximation to
neglect final state interactions for these decays and use the relations 
(\ref{axialc2}) for different intermediate states $M^+$ to
obtain the ratio of the weak form factors between these different states.. 

It is also of interest to look for further tests of the same SU(3) symmetry
in relations between branching ratios of neutral $D$ decays into the final
states which are equally strong and coupled by final state interactions.
In the SU(3) limit these decays satisfy the relations. 
\beq{Dzero1}
{{BR( D^o \rightarrow  K^+ \pi^-)}\over{BR(D^o \rightarrow  
K^-\pi^+) }} =
{{BR[ D^o \rightarrow  K^+  a_1(1260)^-]}\over{BR[D^o \rightarrow  
K^- a_1(1260)^+] }} =
{{BR[ D^o \rightarrow  K^*(892)^+\rho^-]}\over{BR[D^o \rightarrow  
K^*(892)^-\rho^+]}} 
= tan^4 \theta_c 
\eeq 

These relations  should be easily tested and provide useful insight on the
breaking of SU(3) in final state interactions. They involve  no phases and only
branching ratios of decay modes all expected to be comparable to the observed
DCSD $D^o \rightarrow K^+ \pi^-$. 

However, if the SU(3) breaking is really due to the difference between products
of weak axial and strong kaon form factors and vice versa, the relations 
(\ref{Dzero1}) can be expected to be strongly broken and replaced by the 
inequality

\beq{Dzeroin}
{{BR[D^o \rightarrow  
K^- a_1(1260)^+]}\over{BR(D^o \rightarrow  
K^-\pi^+) }} \gg
{{BR[ D^o \rightarrow  K^+  a_1(1260)^-]}\over{
BR( D^o \rightarrow  K^+ \pi^-) }} 
\eeq

\subsection {Relations between $D^+$ and $D_s$ decays} 

We now note an interesting combination of SU(3) relations\cite{plipzh} between
Cabibbo-favored $D^+$ decays and doubly-Cabibbo-forbidden $D_s$ decays and vice
versa. All the obvious SU(3) breaking
effects seem to cancel in this relation and the result is an unambiguous number
which is either right or wrong. 

    Consider the ratio of branching ratios 
 \beq{QQ1a}     
{{BR( D_s \rightarrow  K^+K^+\pi^-)}\over{BR(D_s \rightarrow  
K^+K^-\pi^+)}} \approx O(tan^4 \theta_c)   
\eeq 
 and also the ratio 
 \beq{QQ1b}     
{{BR( D^+ \rightarrow  K^+\pi^+\pi^-)}\over{BR(D^+\rightarrow  
K^-\pi^+\pi^+) }} = {{6.8\pm 1.5 \times 10^{-4}}\over{9.0\pm 0.6\%}}
\approx O(tan^4 \theta_c)   
\eeq
where we have inserted the experimental data for the $D^+$ decays\cite{PDG}. 

Both of these are ratios of a doubly Cabibbo forbidden decay to an allowed
decay and should be of order $tan^4 \theta_c$. The SU(3) transformation
(\ref{QQ3}) takes the two ratios (\ref{QQ1a}) and (\ref{QQ1b}) into the
reciprocals of one another. If strong interaction final state interactions
conserve SU(3) the only SU(3) breaking occurs in the CKM matrix elements and
the product of these two ratios should be EXACTLY $\tan^8 \theta_c$.
 \beq{QQ2}
{{BR( D_s \rightarrow  K^+K^+\pi^-)}\over{BR(D_s \rightarrow K^+K^-\pi^+)}}
\cdot {{BR( D^+ \rightarrow  K^+\pi^+\pi^-)}\over{BR(D^+\rightarrow
K^-\pi^+\pi^+) }} = tan^8 \theta_c 
\eeq 

This includes all SU(3) symmetric final state interactions. Thus if one of
these ratios is enhanced above $tan^4 \theta_c$ as seems to be the case, the
other should be suppressed by the same factor, which is already interesting. 
This relation is seen to have the desirable feature that that most of the
obvious SU(3)-symmetry-breaking factors in the individual SU(3) relations
between the nummerator of one ratio and the denominator of the other seem to
cancel out in this product; e.g. phase space.

	 This result can also be obtained from eq. (3) of ref. \cite{plipzh}
and assuming that the neutral $K\pi$ combinations all come from $K^*$'s. We see
here that the $K^*$ assumption is unnecesary.

Present data\cite{PDG} show
   $ BR(D^+\rightarrow K^+\pi^-\pi^+)/BR(D^+ \rightarrow K^-\pi^+\pi^+) $  is
about 0.65\% or about $3 \times \ tan^4 \theta_c$. This enhancement of the
doubly-Cabbibo-forbidden transition for the $D^+$ decay by a factor of 3 over
the CKM matrix factor is normally explained away by final state interactions.
But if these final state interactions obey SU(3), the relation (\ref{QQ2})
requires the doubly-Cabbibo-forbidden
transition to be suppressed by a factor of 3 for the $D_s$ decay.

   $ BR(D_s\rightarrow K^+K^+\pi^-)/BR(D_s+\rightarrow K^+K^-\pi^+)$  should be
about $(1/3) \times tan^4 \theta_c$ or about 0.07\%.
The effects of the final-state interactions
would differ by about an order of magnitude between $D^+$ and $D_s$ decays.

If on the other hand the $D_s$ decays behave similarly to the $D^+$ decays,
the large violation of SU(3) will need some explanation.
One possibility is always that there may be new physics enhancing the
doubly suppressed decays. These might produce a CP violation which could show
up by looking for a charge asymmetry in the products of above the two ratios;
i.e between the values for $D^+$ and $D_s$ decays and for $D^-$ and $\bar D_s$
decays.

Furthermore, any really large SU(3)-breaking final state interactions
that we don't understand must cast serious doubts on many all analyses and
predictions for heavy-flavor decays which use SU(3) and neglect the possibility
of such final state interactions.

One interesting possibilty might be the enhancement of the ``non-exotic" final
states by the presence of hadronic meson resonances at the $D$ and $D_s$ 
masses. These could explain the enhancement of the doubly-forbidden $D^+$ 
decay and preserve the SU(3) relation by similarly enhancing the allowed $D_s$ 
decay. The doubly-forbidden $D_s$ decay and the allowed $D^+$ decay lead to 
states having exotic quantum numbers which have no resonances. 

An obvious caveat to this analysis is the almost trivial SU(3) breaking arising
from resonances in the final states. But with sufficient data and Dalitz plots
it should be possible to take these resonances into account or look at domains
in the Dalitz plots where they do not appear in order to pinpoint possible 
sources of SU(3) breaking.

In any case the SU(3) relation (\ref{QQ2}) and its possible violations raise
interesting questions which deserve further theoretical and experimental
investigation.

\section{acknowledgments}

It is a pleasure to thank Edmond Berger, Sven Bergmann, Karl Berkelman, John
Cumalat, Yuval  Grossman, Zoltan Ligeti, Yosef Nir and J. G. Smith for helpful
discussions and comments,

{
\tighten

{\begin{figure}[htb]
$$\beginpicture
\setcoordinatesystem units <\tdim,\tdim>
\stpltsmbl
\putrule from -25 -30 to 50 -30
\putrule from -25 -30 to -25 30
\putrule from -25 30 to 50 30
\putrule from 50 -30 to 50 30
\plot -25 -20 -50 -20 /
\plot -25 20 -50 20 /
\plot 50 20 120 40 /
\plot 50 0 120 20 /
\springru 50 -20 *3 /
\plot 120 -20 90 -20 120 -40 /
\put {$b$} [b] at -50 25
\put {$\overline{u}$} [t] at -50 -25
\put {$s$} [l] at 125 40
\put {$\overline{u}$} [l] at 125 20
\put {$q$} [l] at 125 -20
\put {$\overline{q}$} [l] at 125 -40
\put {$\Biggr\}$ $K^-(\vec k)$}[l] at 135 30
\put {$\Biggr\}$ $\eta_1(-\vec k)$} [l] at 135 -30
\put {$G$} [t] at 70 -25
\setplotsymbol ({\large .})
\plot -15 40 60 -40 /
\plot -15 -40 60 40 /
\setshadegrid span <1.5\unitlength>
\hshade -30 -25 50 30 -25 50 /
\linethickness=0pt
\putrule from 0 0 to 0 60
\endpicture$$
\caption{\label{fig-1}} \hfill Forbidden ``gluonic hairpin'' diagram. $G$
denotes any number of gluons. \hfill~
\end{figure}}
 
{\begin{figure}[htb]
$$\beginpicture
\setcoordinatesystem units <\tdim,\tdim>
\stpltsmbl
\putrule from -25 -30 to 50 -30
\putrule from -25 -30 to -25 30
\putrule from -25 30 to 50 30
\putrule from 50 -30 to 50 30
\plot -25 -20 -50 -20 /
\plot -25 20 -50 20 /
\plot 50 20 120 40 /
\plot 50 -20 120 -40 /
\springru 50 0 *3 /
\plot 120 20 90 0 120 -20 /
\put {$b$} [b] at -50 25
\put {$\overline{u}$} [t] at -50 -25
\put {$s$} [l] at 125 40
\put {$\overline{u}$} [l] at 125 20
\put {$u$} [l] at 125 -20
\put {$\overline{u}$} [l] at 125 -40
\put {$\Biggr\}$ $K^-(\vec k)$} [l] at 135 30
\put {$\Biggr\}$  $\eta_u(-\vec k)$} [l] at 135 -30
\put {$G$} [t] at 70 -5
\setshadegrid span <1.5\unitlength>
\hshade -30 -25 50 30 -25 50 /
\linethickness=0pt
\putrule from 0 0 to 0 60
\endpicture$$
\caption{\label{fig-2a}} \hfill Strong $u \bar u$ pair creation. $G$ denotes any number of
gluons. \hfill~ \end{figure}}

{\begin{figure}[htb]
$$\beginpicture
\setcoordinatesystem units <\tdim,\tdim>
\stpltsmbl
\putrule from -25 -30 to 50 -30
\putrule from -25 -30 to -25 30
\putrule from -25 30 to 50 30
\putrule from 50 -30 to 50 30
\plot -25 -20 -50 -20 /
\plot -25 20 -50 20 /
\plot 50 20 120 40 /
\plot 50 -20 120 -40 /
\springru 50 0 *3 /
\plot 120 20 90 0 120 -20 /
\put {$b$} [b] at -50 25
\put {$\overline{u}$} [t] at -50 -25
\put {$s$} [l] at 125 40
\put {$\overline{s}$} [l] at 125 20
\put {$s$} [l] at 125 -20
\put {$\overline{u}$} [l] at 125 -40
\put {$\Biggr\}$  $\eta_s(\vec k)$}[l] at 135 30
\put {$\Biggr\}$  $K^-(-\vec k)$} [l] at 135 -30
\put {$G$} [t] at 70 -5
\setshadegrid span <1.5\unitlength>
\hshade -30 -25 50 30 -25 50 /
\linethickness=0pt
\putrule from 0 0 to 0 60
\endpicture$$
\caption{\label{fig-2b}} \hfill Strong $s \bar s$ pair creation. $G$ denotes any number of
gluons. \hfill~ \end{figure}}

$$ A[\eta_s(\vec k)K^-(-\vec k)] = A[\eta_u(-\vec k)K^-(\vec k)]= P\cdot
A[\eta_u(\vec k)K^-(-\vec k)]$$

{\begin{figure}[htb]
$$\beginpicture
\setcoordinatesystem units <\tdim,\tdim>
\stpltsmbl
\putrule from -25 -30 to 50 -30
\putrule from -25 -30 to -25 30
\putrule from -25 30 to 50 30
\putrule from 50 -30 to 50 30
\plot -25 -20 -50 -20 /
\plot -25 20 -50 20 /
\plot 50 0 120 -20 /
\plot 50 -20 120 -40 /
\photonru 50 20 *3 /
\plot 120 40 90 20 120 20 /
\put {$b$} [b] at -50 25
\put {$\overline{u}$} [t] at -50 -25
\put {$s$} [l] at 125 40
\put {$\overline{u}$} [l] at 125 20
\put {$u$} [l] at 125 -20
\put {$\overline{u}$} [l] at 125 -40
\put {$\Biggr\}$  $K^-(\vec k)$} [l] at 135 30
\put {$\Biggr\}$  $\eta_u(-\vec k)$}  [l] at 135 -30
\put {$W$} [t] at 70 15
\setshadegrid span <1.5\unitlength>
\hshade -30 -25 50 30 -25 50 /
\linethickness=0pt
\putrule from 0 0 to 0 60
\endpicture$$
\caption{\label{fig-3}} \hfill Color favored tree diagram.
 \hfill~ \end{figure}}
 
{\begin{figure}[htb]
$$\beginpicture
\setcoordinatesystem units <\tdim,\tdim>
\stpltsmbl
\putrule from -25 -30 to 50 -30
\putrule from -25 -30 to -25 30
\putrule from -25 30 to 50 30
\putrule from 50 -30 to 50 30
\plot -25 -20 -50 -20 /
\plot -25 20 -50 20 /
\plot 50 20 120 40 /
\plot 50 -20 120 -40 /
\photonru 50 0 *3 /
\plot 120 20 90 0 120 -20 /
\put {$b$} [b] at -50 25
\put {$\overline{u}$} [t] at -50 -25
\put {$u$} [l] at 125 40
\put {$\overline{u}$} [l] at 125 20
\put {$s$} [l] at 125 -20
\put {$\overline{u}$} [l] at 125 -40
\put {$\Biggr\}$ $\eta_u(\vec k)$}  [l] at 135 30
\put {$\Biggr\}$ $K^-(-\vec k)$} [l] at 135 -30
\put {$W$} [t] at 70 -5
\setshadegrid span <1.5\unitlength>
\hshade -30 -25 50 30 -25 50 /
\linethickness=0pt
\putrule from 0 0 to 0 60
\endpicture$$
\caption{\label{fig-4}} \hfill Color suppressed tree diagram.
\hfill~ \end{figure}}

$$ {{A[\eta'(\vec k)K^-(-\vec k)]}\over {A[\eta(\vec k)K^-(-\vec k)]}}= 
{{\langle \eta_u \ket{\eta'}}\over {\langle \eta_u \ket{\eta}}}
$$

\pagebreak
 
  {\begin{figure}[htb]
$$\beginpicture
\setcoordinatesystem units <\tdim,\tdim>
\stpltsmbl
\putrule from -25 -30 to 50 -30
\putrule from -25 -30 to -25 30
\putrule from -25 30 to 50 30
\putrule from 50 -30 to 50 30
\plot -25 -20 -50 -20 /
\plot -25 20 -50 20 /
\plot 50 0 120 -20 /
\plot 50 -20 120 -40 /
\photonru 50 20 *3 /
\plot 120 40 90 20 120 20 /
\put {$c$} [b] at -50 25
\put {$\overline{u}$} [t] at -50 -25
\put {$\overline{d}$} [l] at 125 40
\put {$u$} [l] at 125 20
\put {$s$} [l] at 125 -20
\put {$\overline{u}$} [l] at 125 -40
\put {$\Biggr\}$ Weak   $\rightarrow M^+$} [l] at 135 30
\put {$\Biggr\}$ Strong   $\rightarrow K^-$} [l] at 135 -30
\put {$W$} [t] at 70 15
\setshadegrid span <1.5\unitlength>
\hshade -30 -25 50 30 -25 50 /
\linethickness=0pt
\putrule from 0 0 to 0 60
\endpicture$$
\caption{\label{fig-7}} \hfill Color favored
Cabibbo favored diagram.
 \hfill~ \end{figure}}

  {\begin{figure}[htb]
$$\beginpicture
\setcoordinatesystem units <\tdim,\tdim>
\stpltsmbl
\putrule from -25 -30 to 50 -30
\putrule from -25 -30 to -25 30
\putrule from -25 30 to 50 30
\putrule from 50 -30 to 50 30
\plot -25 -20 -50 -20 /
\plot -25 20 -50 20 /
\plot 50 0 120 -20 /
\plot 50 -20 120 -40 /
\photonru 50 20 *3 /
\plot 120 40 90 20 120 20 /
\put {$c$} [b] at -50 25
\put {$\overline{u}$} [t] at -50 -25
\put {$\overline{s}$} [l] at 125 40
\put {$u$} [l] at 125 20
\put {$d$} [l] at 125 -20
\put {$\overline{u}$} [l] at 125 -40
\put {$\Biggr\}$ Weak   $\rightarrow K^+$} [l] at 135 30
\put {$\Biggr\}$ Strong   $\rightarrow M^-$} [l] at 135 -30
\put {$W$} [t] at 70 15
\setshadegrid span <1.5\unitlength>
\hshade -30 -25 50 30 -25 50 /
\linethickness=0pt
\putrule from 0 0 to 0 60
\endpicture$$
\caption{\label{fig-7}} \hfill Color favored Cabibbo  Doubly-Suppressed
 diagram.
 \hfill~ \end{figure}}

\end{document}